\documentclass[aps,prl,twocolumn,showpacs,superscriptaddress,groupedaddress]{revtex4} 
\usepackage{graphicx} 
\usepackage{dcolumn} 
\usepackage{bm} 
\usepackage{amsmath} 
\usepackage{amssymb} 
\usepackage{color}
\begin{document}
\title{Spiraling eutectic dendrites}
\author{Tam\'as Pusztai,$^1$L\'aszl\'o R\'atkai,$^1$Attila Sz\'all\'as,$^1$ and L\'aszl\'o Gr\'an\'asy}
\affiliation{Institute for Solid State Physics and Optics, Wigner Research Centre for Physics, P.O. Box 49, H-1525 Budapest, Hungary}
\affiliation{BCAST, Brunel University, Uxbridge, Middlesex, UB8 3PH, United Kingdom}
\date{\today}

\begin{abstract}
Eutectic dendrites forming in a model ternary system have been studied using the phase-field theory. The eutectic and one-phase dendrites have similar forms, and the tip radius scales with the interface free energy as for one-phase dendrites. The steady-state eutectic patterns appearing on these two-phase dendrites  include concentric rings, and single- to multiarm spirals, of which the fluctuations choose, a stochastic phenomenon characterized by a peaked probability distribution. The number of spiral arms correlates with tip radius and the kinetic anisotropy. 

\end{abstract}

\pacs{68.70.+w, 81.10.Aj, 81.30.Fb}

\maketitle
Dynamic evolution of spiraling patterns observed in a range of physical, chemical, and biological systems 
including excitable media (such as cellular slime mold \cite{ref1}), Belousov-Zhabotinsky reactions \cite{ref1a}, growth on atomically flat interfaces \cite{ref1b}, binary eutectic systems \cite{ref2}, and more recently, in ternary eutectic systems \cite{ref3} has been exciting the fantasy of researchers for some time. While the details differ, diffusion and phase separation usually play a role. For example, aggregation of starving cells is controlled by propagating spiral waves of a chemo-attractant, often yielding multiarmed spiral patterns \cite{ref1}.  In binary eutectics, spiraling has been associated with specific anisotropy of the solid-solid interface \cite{ref2}, screw dislocations \cite{ref4}, or osmotic flow driven fingering \cite{ref4a}. In turn, the newly discovered spiraling ternary eutectic dendrites emerge from the interplay of two-phase solidification with the Mullins-Sekerka-type diffusional instability caused by the third component \cite{ref3}. This spiraling/helical structure has been identified as of interest for creating chiral metamaterials for optical applications via eutectic self-organization \cite{ref5}. The complex microstructure of some ternary alloys is suspected to originate from eutectic dendrites \cite{ref5a}. Remarkably, multiarm spiraling has been reported experimentally in  excitable media \cite{ref1}, in binary eutectics \cite{ref2}, and in Liesegang reactions \cite{ref5b}, and theoretically in the FitzHugh-Nagumo model, in which the multiarm spirals form due to the attraction of single spirals  \cite{ref6}. It is yet unclear how general this behavior is, in particular whether multiarm spiraling is possible for ternary eutectic dendrites, and what governs the number of spiraling eutectic arms.  

In this paper, we show that a minimal phase-field model of ternary freezing is able to describe the spiraling ternary eutectic dendrites, and perform a detailed numerical study of this exotic growth mode. We demonstrate that the multiarm eutectic spiral patterns are robust, so they should be experimentally accessible, and that analogously to the findings for Liesegang reactions \cite{ref5b}, the number of spirals results from an interplay of  stochastic effects and the competition of nonlinear modes.

The free energy of a minimal ternary generalization of the binary phase-field model (see e.g. Ref. \cite{ref7a}) reads as 
\begin{eqnarray}
\begin{split}
F[\phi,\mathbf{c}] = \int \left[ \frac{\epsilon_{\phi}^2}{2} (\nabla\phi)^2 + w g(\phi) +  (1-p(\phi))f_l(\textbf{c}) + \right.\\
\left.   + p(\phi) \left( f_s(\textbf{c}) + \frac{\epsilon_c^2}{2} \sum_{i=1}^{3} (\nabla {c_i})^2 \right) \right] \, dV,
\end{split}
\end{eqnarray}
where $\phi \in [0, 1]$, $\epsilon_{\phi}, \epsilon_c,$ and $w$ are constants, for the $g(\phi)$ and $p(\phi)$ functions see Ref. \cite{ref7a}, whereas  $\mathbf{c}=(c_1, c_2, c_3)$, and the bulk liquid and solid phases are regarded as ternary ideal and regular solutions:
\begin{eqnarray}
f_{l,s}(\mathbf{c}) = \sum_{i=1}^3 c_i \left[f^{l,s}_{i} + \log c_i \right] + \frac{1}{2}\sum_{i,j, i \ne j} \Omega^{l,s}_{ij} c_i c_j.
\end{eqnarray}
The equations of motion (EOMs) have been derived variationally, yielding
\begin{eqnarray}
\begin{split}
\dot{\phi} = M_{\phi} \left[ \epsilon_{\phi}^2 \nabla^2 \phi - w g^\prime(\phi) + \right. \\
\left. + p^{\prime}(\phi)(f_l(\mathbf{c}) - f_s(\mathbf{c})) - p^{\prime}(\phi) \epsilon_c^2 \sum_{i=1}^{3} (\nabla {c_i})^2 \right]
\end{split}
\end{eqnarray}
for the phase field, and 
\begin{eqnarray}
\dot{c_i} = \sum_{j=1}^{3}\nabla \cdot \left[(1-p(\phi))M^c_{i,j}\left(\nabla\frac{\delta{F}}{\delta{c_j}}\right)\right],
\end{eqnarray}
for the concentration fields, where the $\sum_i c_i = 1$ constraint is automatically satisfied by our choice of the specific values of 1 and $-0.5$ for the diagonal and off-diagonal elements of the 3 $\times$ 3 mobility matrix, $\mathbf{M^{c}}$. 

The dimensionless form of these EOMs has been solved parallel by finite differencing and explicit time stepping on a 3D grid, on a cluster of computers. Most of our simulations have been carried out in a directional solidification configuration. A temperature gradient was implemented by making the solid free energy temperature dependent as $f_{s,i,\tilde{z}} = f_{s,i}^{(0)}-\tilde{z}\left(\partial{f_{s,i}} / \partial{\tilde{z}}\right)$, where $\tilde{z}$ is the coordinate along the direction of $\tilde{v}_p$ sample pulling. Sample pulling has been modeled by shifting the contents of the arrays $\phi$ and $c_i$ by one voxel back in the $\tilde{z}$ direction in each $[(d\tilde{x}/d\tilde{t})/\tilde{v}_p]$\textsuperscript{th} time step, with boundary conditions $\phi=0$ and $\mathbf{c} = \mathbf{c_0}$ on the high $T$ and no flux boundary conditions on the low $T$ side of the sample.  To enable large enough simulation boxes in the direction of pulling ($\tilde{z}$), only Eqs. ({4}) have been solved far ahead of the solidification front, where $\phi$ is sufficiently small ($< 10^{-8}$). Since the anisotropy of the solid-liquid interface free energy is weak for metals and the transparent system used in Ref. \cite{ref3}, we have considered only kinetic anisotropy (of cubic symmetry, see \cite{ref7b}).
Solidification has been started by including a slab of solid of length $\tilde{L}_z (2/9)$, where $\tilde{L}_z = N_z d\tilde{x}$ is the length in direction $\tilde{z}$ \cite{ref8} with a small hump at the center, whereas the initial composition of the solid has been $\langle c_1 \rangle = \langle c_2 \rangle = 0.455$ and $c_3 = 0.09$ realized by a random transversal ($\tilde{x}-\tilde{y}$ plane) distribution of the two solid phases. We have opted for this starting condition, because simulations that follow the formation of the two-phase dendrites from fluctuation-induced emergence of the Mullins-Sekerka instability of a flat interface are prohibitively time consuming. We have, however, demonstrated the formation of surface undulations increasing with time, and  found that the lower and upper unstable wavelengths 
are $\sim 60$ and  $\sim 320$, with the fastest growing wavelength being around $\sim 120$. The parallel computations of this study would have taken more than 600 years on a single CPU core.

\begin{figure}
\includegraphics[width=0.8\linewidth]{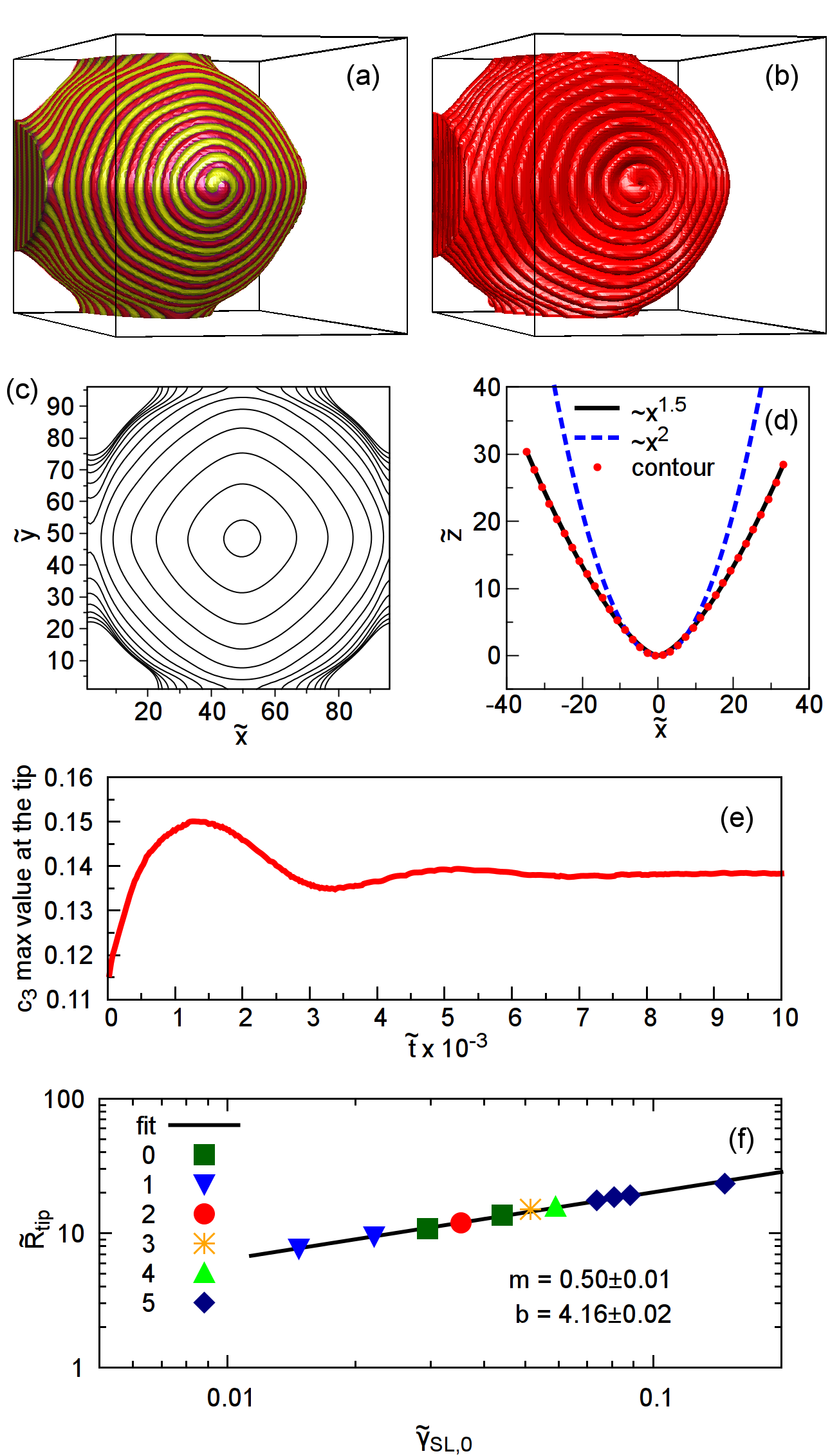}
\caption{(color online) Two-phase spiraling dendrite grown under conditions given in Ref. \cite{ref8}: (a) spiraling motif on the surface; (b) the helical structure formed by one of the solid phases, (c) contour lines showing the transverse sections at $10 d\tilde{x}$ distances; (d) longitudinal section (dots), the best fit parabola (dashed),  and the curve $\tilde{z} = \tilde{z}_{max}-|\tilde{x}|^{\nu}$ fitted to it (solid line); (e) maximum of $c_3$ at the tip vs time; (f) tip radius vs. solid-liquid interface energy. 
} 
\label{fig:single}                                                              
\end{figure}

In our study, first we have explored the parameter space defined by composition, temperature gradient, pulling velocity, interfacial free energy, and kinetic anisotropy, and optimized the conditions for growing two-phase steady-state dendritic structures. Typical conditions for such dendrites are summarized in Ref. \cite{ref8}. For low pulling velocities ($\tilde{v}_p < 0.03$) one finds a planar front with lamellar pattern, whereas at high enough pulling velocities ($\tilde{v}_p > 0.3$), solidification takes place without apparent partitioning, though the dendritic structure is yet preserved. At even higher pulling velocities ($\tilde{v}_p > 0.8$) partitionless growth with a flat interface has been found. A typical two-phase dendrite [see Figs. 1(a) and 1(b)] has a rounded square-like transverse section in the $\tilde{x}-\tilde{y}$ plane [Fig. 1(c)]; whereas in the fin directions (e.g., $\tilde{x}-\tilde{z}$), the longitudinal profile can be fitted with $\tilde{z} = \tilde{z}_{max}-|\tilde{x}|^{\nu}$, where $\tilde{z}_{max}$ is the tip position, $\tilde{x}$ is the distance from the axis of the dendrite, and $\nu$ is $\sim 1.49 \pm 0.05$ [Fig. 1(d)], somewhat lower than the $\nu = 1.67$ found experimentally for  single-phase xenon dendrites \cite{ref9}. (The perimeter of the dendrite has been determined by the contour line $\phi = 0.5$.) This steady-state shape has been achieved after a transient composed of decaying oscillations of the tip-radius, tip-temperature, and the maximum of $c_3$ at the tip [Fig. 1(e)]. To test further, how far the two-phase dendrites resemble the single-phase dendrites, we have varied the magnitude of the solid-liquid interface free energy ($\tilde{\gamma}_{SL}$), via changing the free energy of the single component solid-liquid interface ($\tilde{\gamma}_{SL,0}$) and evaluated the tip radius in the fin direction ($\tilde{R}_{tip}$). The results indicate $\tilde{R}_{tip} \propto \tilde{\gamma}_{SL,0}^{0.50 \pm 0.01}$ 
[see Fig. 1(f)], which is in a good agreement with $\tilde{R}_{tip} \propto \tilde{\gamma}_{SL}^{1/2}$ derived theoretically for single-phase dendrites \cite{ref10}, and may indicate e.g. a negligible chemical contribution to $\tilde{\gamma}_{SL}$. Apparently, the shape of the two-phase dendrite is independent from the eutectic pattern forming the solid dendrite: target patterns, single- and multiple spiraling motifs do coexist on the same $\tilde{R}_{tip}$ vs. $\tilde{\gamma}_{SL,0}$ curve [see Figs. 1(f) and 2]. 

\begin{figure}
\includegraphics[width=0.9\linewidth]{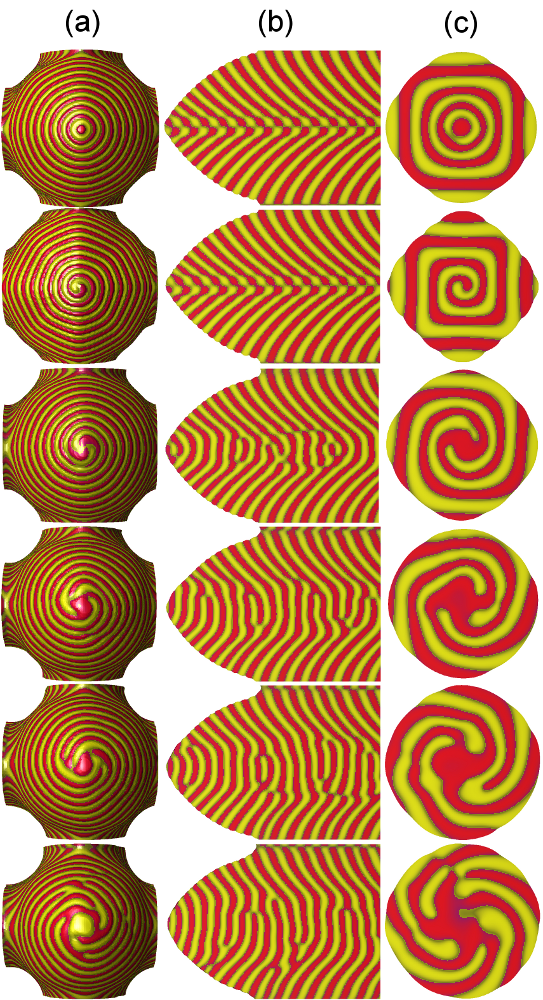}
\caption{(color online) Eutectic patterns of two-phase dendrites. (a) Front view; (b) longitudinal  and (c) transverse sections. From top to bottom, $\tilde{\gamma}_{SL,0} = 0.0295,$ $0.0147,$ $0.0354,$ $0.0516,$ $0.0589,$ and $0.0810$, respectively. 
The disorder in the tip region increases with increasing interfacial free energy.
}
\label{fig:multi}                                                              
\end{figure}                                                                     

The target pattern advances via alternating nucleation of the two solid phases, a mode expected to disappear at small undercoolings. It is more frequent for lower interface free energies, and becomes rare for $\tilde{\gamma}_{SL,0} > 0.1$. Besides the target pattern, a number of steady state ``spiraling" modes have been observed that display one to five arms [Fig. 2]. (On the surface of the dendrite spirals are realized by helical structures forming in the volume.)  Owing to evident geometrical constraints, the steepness of the spirals increases with the number of the arms. The longitudinal sections are fairly similar for all modes, although weak systematic differences are observed. More characteristic are the front views and the transverse sections: The individual modes (number of spirals) can clearly be distinguished [Figs. 2(a) and 2(c)]. We also find that once in the appropriate parameter domain, the spiraling two-phase dendrites are quite robust. We note that in the experiments, which were performed at low undercoolings, only the single-spiral mode has been observed so far \cite{ref3}. The large number of spiral arms seen here probably follows from the large relative undercooling \cite{ref8} used in our simulations.

The larger the tip radius, the larger is the number of spiraling arms [Figs. 1(f) and 2]. A closer inspection of the tip region reveals that no nucleation is needed for the single spiral mode, where the two-phase spirals originate from a rotating ``yin-yang" like motif at the tip of the eutectic dendrite. The modes with larger number of spirals become increasingly more complex, displaying alternating phase appearance at the tip. It is difficult to decide whether heterogeneous nucleation or growth around the phase occupying the tip is the mechanism by which the phases invade the tip. With the exception of the target pattern, where cones of one of the solid phases are not connected with other cones of the same phase, in spiraling modes the individual one-phase regions are interconnected with all volumes of that phase [Fig. 1(a)]. 

\begin{figure}
\includegraphics[width=0.85\linewidth]{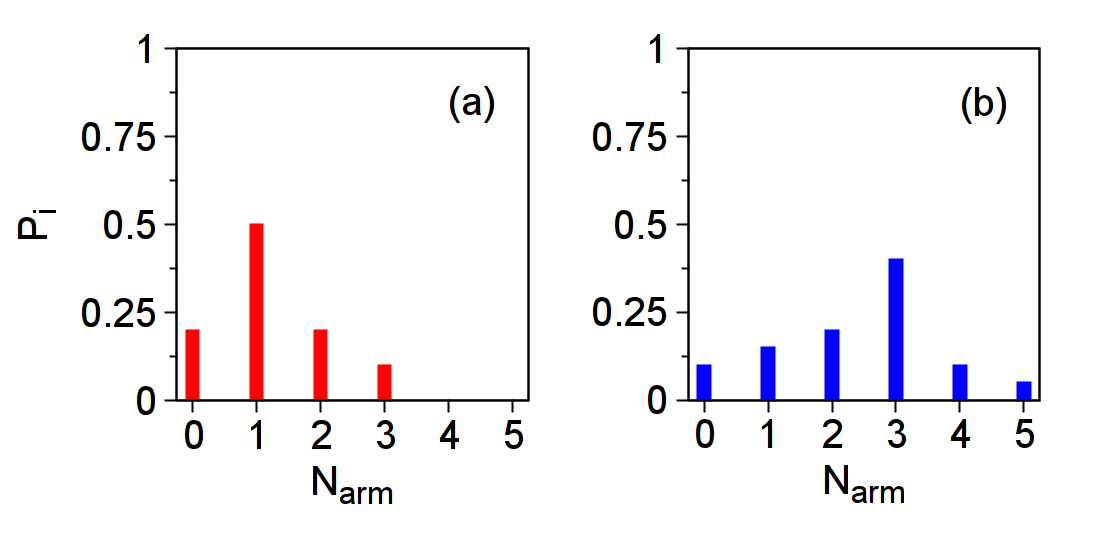}
\caption{(color online) Probability distribution of the steady state dendritic patterns from 20 different random initial two-phase patterns. (a) $\tilde{\gamma}_{SL,0} = 0.0354$; (b) $\tilde{\gamma}_{SL,0} = 0.0589$.  
} 
\label{fig:statistics}                                                              
\end{figure}                                                                     

Although the number of spiraling arms ($N_{arm}$) tends to increase with the solid-liquid interfacial free energy ($\tilde{\gamma}_{SL,0}$) [Figs. 1(f) and 2], the steady-state pattern appearing after the transient depends also on the initial random distribution of the two solid phases: Different steady state patterns are obtained starting from different (random) initial patterns. For example, at $\tilde{\gamma}_{SL,0} = 0.0354$, these patterns include the target pattern, and single- to triple spirals (Fig. 3), showing a multiplicity of steady-state solutions for nominally the same conditions of which random initialization (representing here the cumulative effect of preceding compositional fluctuations) chooses. In other words, the thermal fluctuations decide which steady-state solutions are accessible for the system under a given set of operating parameters. Indeed, we have observed a similar stochastic behavior, when initiating growth with a chemically homogeneous solid slab, and adding noise representing fluctuations to the EOMs (a study inspired by Ref. \cite{ref5b}). These features closely resemble the helical Liesegang patterns, where the thermal fluctuations determine, which of the competing modes (helical, double helical, or non-helical) is realized \cite{ref5b}. These similarities raise the possibility of a universal behavior for a class of multiarm spiral systems. Whether such a stochastic behavior prevails in other multiarm spiral systems requires further investigations.

\begin{figure}
\includegraphics[width=0.8\linewidth]{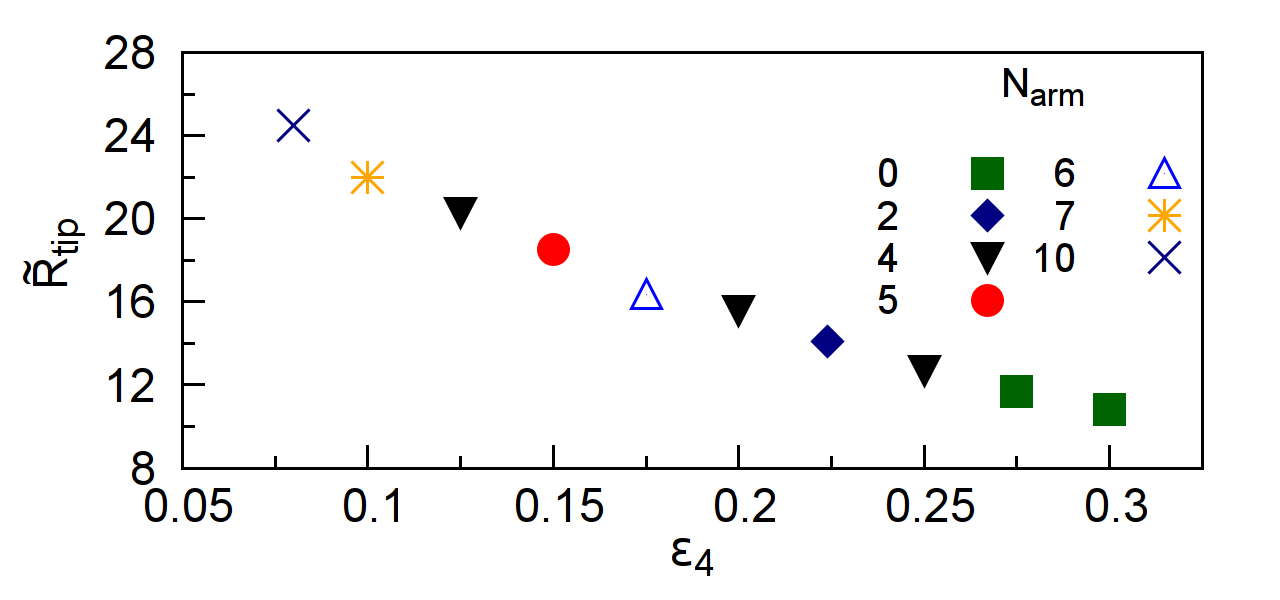}
\caption{(color online) Tip radius vs. kinetic anisotropy at $\tilde{\gamma}_{SL,0}= 0.0295$. The number of spiral arms tends to increase with decreasing anisotropy, although with some scattering.  
} 
\label{fig:radius}                                                              
\end{figure}                                                                     

Next, we investigate, how the kinetic anisotropy influences the number of spirals. We find that with decreasing anisotropy the tip radius increases followed by the number of the spirals, which however shows some stochastic scattering (Fig. 4). We note, furthermore, that the exponent $\nu$ describing the shape of the dendrite tip changes from $\sim 1.49 \pm 0.1$ to $\sim 2.1 \pm 0.1$, varying between roughly the experimental value for xenon (1.67) and the rotational paraboloid (2.0) expected for isotropic case. This is combined with a change of the transverse section from a square of rounded corners to a circle. 

\begin{figure}
\includegraphics[width=0.8\linewidth]{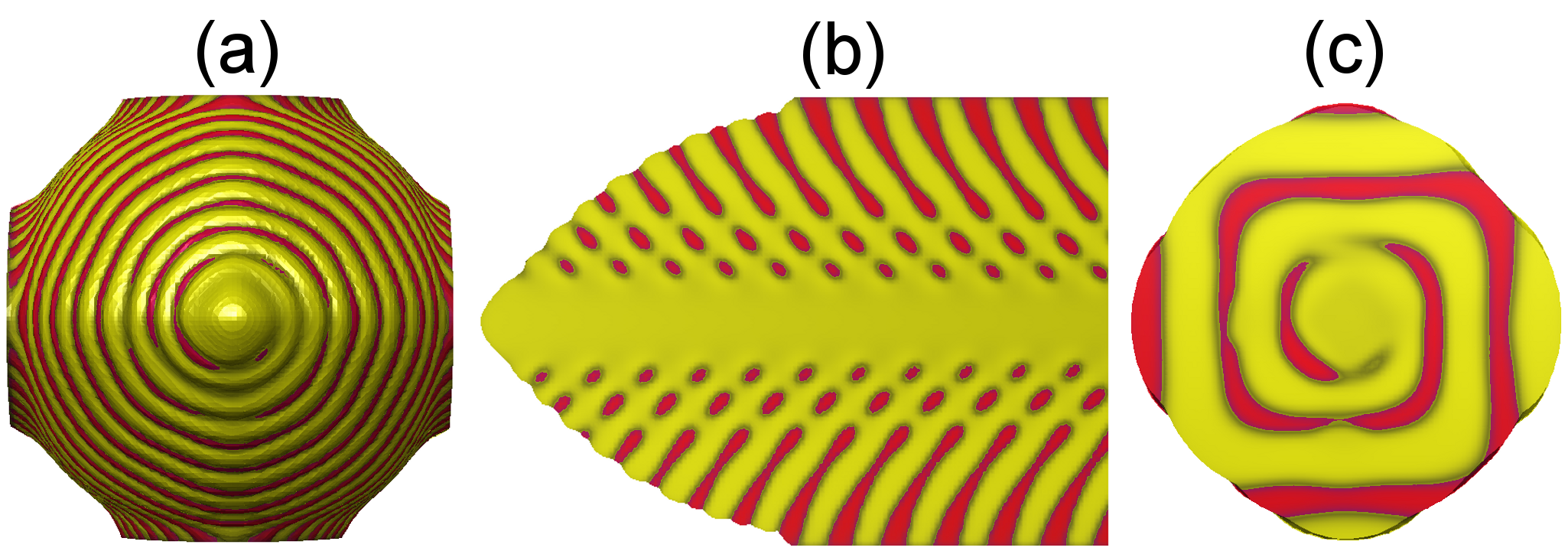}
\caption{(color online) Spiraling dendrite formed at $c_1 = 0.355$ while gradually changing  $c_1/c_2$ at $c_3 = 0.09$ in the incoming liquid. (a) Front view; (b) longitudinal  and (c) transverse sections.  Note the majority phase channel at the center.   
} 
\label{fig:off-eut}                                                              
\end{figure}                                                                     

Finally, we explore how the two-phase pattern varies for off-eutectic compositions. Eutectic dendrites have been seen to form only close to the eutectic composition. With a slow change of the liquid composition, one can move yet away from the eutectic composition, while retaining the spiraling structure (Fig. 5). Beyond a critical deviation from the eutectic composition, the majority phase forms a channel at the centerline of the two-phase dendrite, a feature apparent in the experimental observations \cite{ref3}. Remarkably, such patterns exist in the steady state, raising the possibility that their apparent lack of formation from random eutectic pattern, is only due to a long relaxation time, inaccessible for our simulations. 

Summarizing, we have shown that the ternary phase-field model naturally incorporates the spiral eutectic dendrites, and that such two-phase growth forms emerge between the domains of lamellar eutectic patterns and solute trapping. The two-phase dendrites behave analogously to their single-phase counterparts, whereas the underlying eutectic pattern has little influence on the shape. A number of eutectic growth modes compete, including the target pattern, and single to multiple spirals, of which thermal fluctuations choose. The number of spiral arms tends to increase with the tip radius or interface free energy, and decrease with the kinetic anisotropy. These findings are expected to instigate further experimental/theoretical studies on multiarm spiral systems and their stochastic nature. 

\begin{acknowledgments} This work has been supported by the EU FP7 
projects ``ENSEMBLE" (Grant Agreement NMP4-SL-2008-213669) and 
``EXOMET" (contract No. NMP-LA-2012-280421, co-founded by ESA). We 
thank Mathis Plapp and Zolt\'an R\'acz for the enlightening discussions.
\end{acknowledgments}


\begin{thebibliography}{99}

\bibitem{ref1} F. Siegert and C. J. Weijer, Curr. Biol. {\bf 5}, 937 (1995); J. Rietdorf, F. Siegert, and 
C. J. Weijer, Dev. Biol. {\bf 177}, 427 (1996).

\bibitem{ref1a} R. V. Suganthi {\it et al.}, 
     J. Mater. Sci.: Mater. Med. {\bf 20}, S131 (2009). 

\bibitem{ref1b} C. Klemenz, J. Cryst. Growth {\bf 187} 221 (1998); A. Karma and M. Plapp, Phys. Rev. Lett. {\bf 81}, 4444 (1998).

\bibitem{ref2} R. L. Fullman and D. L. Wood, Acta Metall. {\bf 2}, 188 (1954); H. Y. Liu and H. Jones, Acta Metall. Mater. {\bf 40}, 229 (1992). 

\bibitem{ref3} S. Akamatsu {\it et al.}, 
     Phys. Rev. Lett. {\bf 104}, 056101 (2010). 

\bibitem{ref4} Yu. N. Taran, V. I. Mazur, and P. V. Terent'eva, Izv. Akad. Nauk. SSSR, Metally, No. 1, 25 (1976).

\bibitem{ref4a} G. Tegze and G. I. T\'oth, Acta Mater. {\bf  60}, 1689 (2012).


\bibitem{ref5} D. A. Pawlak {\it et al.}, 
     EU FP7 Collaborative Project ``ENSEMBLE", CF-FP 213669, Grant Agreement, Annex I. 

\bibitem{ref5a} S. A. Souza {\it et al.}, 
     J. Alloys Compounds {\bf 402}, 156 (2005).

\bibitem{ref5b} S. Thomas, I. Lagzi, F. Moln\'ar, Z. R\'acz, editorially approved for publication in Phys. Rev. Lett. (2013).


\bibitem{ref6} B. Vasiev, F. Siegert, and C. Weijer, Phys. Rev. Lett. {\bf 78}, 2489 (1997).

\bibitem{ref7a}  L. Gr\'an\'asy {\it et al.}, Phys. Rev. Lett. {\bf 88}, 206105 (2002).

\bibitem{ref7b} The expression for anisotropy:
$\epsilon (\textbf{n}) = (1-3\epsilon_4)$
$\left[1 - \frac{4\epsilon_4}{1-3\epsilon_4}\right]$ $(n_x^4 +n_y^4 +n_z^4)$.
Here $\textbf{n} = (n_x, n_y, n_z)$ is the surface normal and $\epsilon_4$ the anisotropy parameter. 

\bibitem{ref8} Conditions for the simulation shown in Figs. 1(a) - 1(e). (These are used in other simulations unless stated otherwise): 
Time and spatial steps: $d\tilde{t} = 0.0025$ and $d\tilde{x} = 1.0$.
Grid: $N_x \times N_y \times N_z$ = $96 \times 96 \times 612$.
Size of simulation box: $\tilde{L}_i = N_i \cdot d\tilde{x}$, where $i = x, y,$ or $z$.
Composition: $c_1 = c_2=0.455$, and $c_3 = 0.09$. 
Parameters of free energy densities: $f_{l,i}=0$; $f_{s,i}^{(0)}=-0.9640$ and $\partial{f_{s,i}}/\partial{\tilde{z}}=3.677 \times 10^{-4}$. At $\tilde{t} = 9000$ (steady state), this yields $f_{s,i} = -0.9140$ at the dendrite tip, corresponding to a rather substantial relative undercooling of $\Delta \tilde{T}_r = (\tilde{T}_L-\tilde{T})/(\tilde{T}_L-\tilde{T}_S) \approx 0.78$, where $\tilde{T}_L$ and $\tilde{T}_S$ are the dimensionless liquidus and solidus temperatures corresponding to the conditions at the dendrite tip.
Pulling velocity: $\tilde{v}_p=0.2$, whereas $\Omega_{i,j}^l=0$,  $\Omega^s_{1,2} = 3.05$,  $\Omega^s_{2,3} = \Omega^s_{3,1} = 3.0$. $\tilde{M}_{\phi} = 1.0.$ 
Solid-liquid interface free energy (isotropic): $\tilde{\gamma}_{SL,0} = 0.0147$. 
Kinetic anisotropy: $\epsilon_4 = 0.3$. $\epsilon_{\phi}^2 = 0.75, \epsilon_{c}^2 = 0.4,$ and $w = 0.0469$.

\bibitem{ref9} H.M. Singer and J.H. Bilgram, Phys. Rev. Lett. {\bf 75}, 3898 (1995).

\bibitem{ref10} See e.g., W. Kurz and D. J. Fisher, {\it Fundamentals of Solidification} (Trans. Tech, Lausanne, 1989).

\end{thebibliography}
\end{document}